\documentclass[10pt,letterpaper]{article}
\usepackage[top=0.85in,left=2.75in,footskip=0.75in,marginparwidth=2in]{geometry}

\usepackage[utf8]{inputenc}

\usepackage{cite}

\usepackage{nameref,hyperref}

\usepackage[right]{lineno}

\usepackage{microtype}
\DisableLigatures[f]{encoding = *, family = * }

\raggedright
\usepackage{changepage}

\usepackage[aboveskip=1pt,labelfont=bf,labelsep=period,singlelinecheck=off]{caption}

\usepackage[parfill]{parskip}

\makeatletter
\renewcommand{\@biblabel}[1]{\quad#1.}
\makeatother

\usepackage{lastpage,fancyhdr,graphicx}
\usepackage{epstopdf}
\fancyheadoffset[L]{2.25in}
\fancyfootoffset[L]{2.25in}

\usepackage{listings}
\usepackage{color}

\definecolor{lightgray}{rgb}{.9,.9,.9}
\definecolor{darkgray}{rgb}{.4,.4,.4}
\definecolor{purple}{rgb}{0.65, 0.12, 0.82}

\lstdefinelanguage{JavaScript}{
  keywords={typeof, new, true, false, catch, function, return, null, catch, switch, var, if, in, while, do, else, case, break},
  keywordstyle=\color{blue}\bfseries,
  ndkeywords={class, export, boolean, throw, implements, import, this},
  ndkeywordstyle=\color{darkgray}\bfseries,
  identifierstyle=\color{black},
  sensitive=false,
  comment=[l]{//},
  morecomment=[s]{/*}{*/},
  commentstyle=\color{purple}\ttfamily,
  stringstyle=\color{red}\ttfamily,
  morestring=[b]',
  morestring=[b]"
}

\usepackage{color}

\definecolor{Gray}{gray}{.25}

\usepackage{graphicx}

\usepackage{sidecap}

\usepackage{wrapfig}
\usepackage[pscoord]{eso-pic}
\usepackage[fulladjust]{marginnote}
\reversemarginpar

\usepackage[symbol]{footmisc}

\begin{document}
\vspace*{0.35in}

\begin{flushleft}
{\Large
\textbf\newline{pdbmine: A Node.js API for the RCSB Protein Data Bank (PDB)}
}
\newline
\\
Newlyn N. Joseph\textsuperscript{1*},
Raktim N. Roy\textsuperscript{1},
Thomas A. Steitz\footnote[2]{Deceased 9 October 2018},
\\
\bigskip
\bf{1} Department of Biophysics and Biochemistry, Yale University
\\
\bigskip
* newlyn.joseph@yale.edu

\end{flushleft}

\section*{Abstract}
\textbf{Summary:} The advent of Web-based tools that assist in the analysis and visualization of macromolecules require application programming interfaces (APIs) designed for modern web frameworks. To this end, we have developed a Node.js module \texttt{pdbmine} that allows any user to generate faster data-request queries  to the RCSB Protein Data Bank (PDB). This JavaScript API acts as a layer over the XML-based RCSB PDB RESTful API. The relatively simple nature of the function calls within this module allows the user to easily implement and integrate \texttt{pdbmine} into larger Node.js web applications.\\
\textbf{Availability:} This module can be installed via the Node Package Manager (NPM) at \url{https://www.npmjs.com/package/pdbmine/}, and is hosted on GitHub under the open-source MIT license at \url{https://github.com/nnj1/pdbmine/}. Relevant documentation is detailed at \url{https://nnj1.github.io/pdbmine/}\\


\section*{Introduction}
The RSCB Protein Data Bank (PDB) contains a plethora of structural biology information, including X-ray diffraction structures, CryoEM models and NMR ensembles \cite{berman2000protein}. These structures are widely accessed and used for structural analysis and computational work. The PDB website provides various in-browser features that enable researchers to visualize, manipulate, and inspect structures, as well as obtain meta-data pertaining to a structure \cite{burley2018rcsb}.

The PDB database can be envisioned as a representative of the expanding initiative to move scientific tools and resources into web environments, where the prerequisite of stringent hardware and softwares can be easily bypassed through a mere web connection\cite{gilbert2004bioinformatics}.

Web applications that offer scientific tools and analytical functionalities usually require access to the PDB. While the PDB database does provide a RESTful API for making data requests, an additional layer of abstraction in the form of a language-specific API will assist researchers and developers in incorporating PDB data into their applications. This article introduces a Node.js API for querying and requesting data from the PDB.

Modern web applications are increasingly being developed using Node.js, a server-side JavaScript run-time environment that leverages Google's V8 JavaScript engine \cite{tilkov2010node}. Advantages of using this framework include easy scalability, seamless integration of client-side and server-side code, and modular development practices that utilize of the node package manager registry. \texttt{pdbmine} aims to address the lack of Node.js specific PDB querying APIs. This API essentially acts as a layer of abstraction over the XML query requests that the PDB RESTful API generally accepts.

\section*{Usage}

This module can be easily acquired using the \texttt{npm} command line tool and invoking \texttt{npm install pdbmine}. Once the module is added as a dependency to a project, it can be referenced in scripts and used to make various PDB queries. The core functions of this library are described in Table 1. 

\begin{table}[!ht]
\begin{adjustwidth}{-1.5in}{0in} 
\centering
\caption{{\bf List of the functions provided by this API.} Further details are available in the full documentation which is available on the GitHub repository.}
\begin{tabular}{|p{4cm}|p{13cm}|}
\hline
\bf{Function} & \bf {Details}\\ \hline
\texttt{get\_all\_ids(cb)} & Returns a list of strings containing every PDB ID in the database as an argument to the callback. \\ \hline
\texttt{query(query\_string, cb)} & Accepts a query string argument and returns a list of PDB IDs pertinent to the search query as an argument to the callback,    \\ \hline 
 \texttt{describe\_pdb(ids, params, cb)} & Accepts a list of PDB IDs, along with a list of parameters to return for each ID. Returns a list of JSON documents relevant to the provided PDB IDs as an argument to the callback.  \\  \hline
 \texttt{download(id, format, cb)} & Accepts PDB ID along with a string specifying the format (\texttt{.pdb} or \texttt{.cif}), and returns a a string representation of the file as an argument to the callback. \\   \hline

\end{tabular}
\label{tab1}
\end{adjustwidth}
\end{table}

Common to all methods is the callback parameter, which is a function that is invoked after the asynchronous HTTP request returns the relevant data. However, after the ECMAScript 6 Standard introduced the concept of JavaScript Promises, support for Promises was added to the above functions \cite{wirfs2015ecmascript}. Usage details are described in the module documentation.

The \texttt{query} function returns a array of JSON documents, given that JSON is the most natural JavaScript data structure for document-style data. Arrays in JavaScript can be filtered using custom functions. Hence, the JSON documents returned by \texttt{query} be further filtered based on conditions on the comprising element values. An example of such usage is depicted in the code block below:

\begin{lstlisting}[language=JavaScript]
miner.query("ribozyme")
	 .then(results => miner.describe_pdb(results, ["macromoleculeType"]))
	 .then(descriptions => descriptions.filter(obj => obj.macromoleculeType == "RNA"))
	 .then(rnas => console.log(rnas));

\end{lstlisting}

The output is a list of structures pertaining to the ``ribozyme'' query that consist of RNAs. Additional methods are in the process of development and any contributions to the Git repository for the project are welcomed. 

\newpage 

\subsection*{Example Case}

\begin{wrapfigure}[12]{l}{90mm}
\includegraphics[width=90mm]{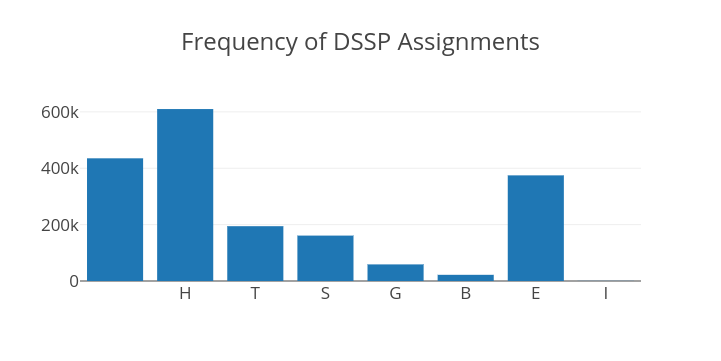}
\captionsetup{labelformat=empty} 
\caption{} 
\label{fig1} 
\end{wrapfigure}

\marginpar{
\vspace{1.5cm} 
\color{Gray} 
\textbf{Figure \ref{fig1}. Example of usage.} 
A plot.ly graph displaying the frequency of various DSSP assignments for structures corresponding to the query ``transmembrane protein.''
}

Figure 1 displays an example usage case for this module. The PDB was queried for the terms ``transmembrane protein,'' and the \texttt{kabschSander} parameter was requested. One of eight characters, corresponding to a secondary structural element, is assigned to each residue in the primary structure \cite{kabsch1983dictionary}.
The resulting 1,857,175 total assignments corresponding to the matching PDB entries were plotted using the external \texttt{plot.ly} Node.js module \cite{plotly}. This serves as one simple example for the kind of data wrangling that this module aims to simplify.

\section*{Conclusion}

We have introduced an API for Node.js that serves as a wrapper around the RESTful RSCB PDB API, for the purpose of simplifying the task of PDB querying in modern web applications. This tool can fit well into larger web-based bioinformatics pipelines that may make use of PDB/mmCIF parsers and/or external visualization modules.

The latest release of the code is available on the npm registry and issues, along with development progress, can be tracked on the GitHub repository.

\section*{Funding}

This work has been supported by the Howard Hughes Medical Institute and the National Institutes of Health [GM022778 to T.A.S.].


\nolinenumbers

\bibliography{library}

\bibliographystyle{abbrv}

\end{document}